\documentclass [12pt]{article}
\usepackage[round]{natbib}

\usepackage {graphicx}
\usepackage{epsfig,amsmath,latexsym,amssymb}
\usepackage{bm}

\usepackage{booktabs}
\usepackage{multirow}
\newtheorem{algorithm_new}{Algorithm}[section]

\begin{document}

\title{\vspace{-3cm}Valuation of  Variable Annuities with Guaranteed Minimum Withdrawal and Death Benefits via Stochastic Control Optimization}

\author{Xiaolin Luo$^{1,\ast}$ and Pavel V.~Shevchenko$^{2}$}

\date{\footnotesize{Draft, 1st version 20 November 2014, this version 10 February 2015 }}

\maketitle

\begin{center}
\footnotesize { \textit{$^{1}$ The Commonwealth Scientific and Industrial Research Organisation, Australia; e-mail: Xiaolin.Luo@csiro.au \\
$^{2}$ The Commonwealth Scientific and Industrial Research Organisation, Australia;
e-mail: Pavel.Shevchenko@csiro.au  \\
$^*$ Corresponding author} }
\end{center}

\begin{abstract}
\noindent
  In this paper we present a numerical valuation of variable
  annuities
 with combined Guaranteed Minimum Withdrawal Benefit (GMWB) and
Guaranteed Minimum Death Benefit (GMDB) under
optimal policyholder behavior solved as an optimal stochastic control problem. This product simultaneously
deals with financial risk,  mortality risk and human behavior. We assume that market is complete in financial risk and mortality risk is completely diversified by selling enough policies and thus the annuity price can be expressed as appropriate expectation.  The computing engine employed to solve the optimal stochastic control problem is
based on a robust and efficient Gauss-Hermite quadrature method with cubic spline.
We present results for three different types of death benefit and show that, under the optimal policyholder behavior,  adding the premium for the death benefit on top of the GMWB can be problematic for contracts with long maturities if the
   continuous fee structure is kept, which is ordinarily assumed for a GMWB
   contract. In fact for some long maturities it can be shown that
   the fee cannot be charged as any proportion of the account value --  there is no solution to match the initial premium
   with the fair annuity price. On the other
  hand, the extra fee due to adding the death benefit can be
  charged upfront or in periodic instalment of fixed amount, and it is cheaper than buying a separate life insurance.


\vspace{0.5cm} \noindent \textbf{Keywords:} \emph{Variable Annuity,
Optimal Stochastic Control, Guaranteed Minimum Withdrawal Benefit,
Guaranteed Minimum Death Benefit, Mortality Risk}
\end{abstract}

\pagebreak

\section{Introduction}
\label{sec:introduction}

A variable annuity is a fund-linked insurance contract including a
variety  of financial options on the policy account value; see e.g.
\cite{Smith1982lifeinsurance} and \cite{Walden1985wholelife}.  A
recent description of the main features of variable annuity products
and the development of their market  can be found in
\cite{Ledlie2008variableannuities} and \cite{Kalberer2009}.
 The main features of variable annuities are
represented by a  variety of possible guarantees, which are briefly
referred to as GMxB - Guaranteed Minimum `x' Benefit, where `x'
stands for accumulation (A), death (D), income (I) or withdrawal
(W). All GMxBs provide a protection of the policyholder's  account:
GMAB during the accumulation phase and  GMDB in case of early death,
GMIB and GMWB after retirement, in particular in the face of high
longevity. In this study we focus on Guaranteed Minimum Withdrawal
Benefit (GMWB) in combination with Guaranteed Minimum Death Benefit (GMDB), overall referred to as GMWDB.

 A variable annuity contract with
GMWB promises to return the entire initial investment through cash
withdrawals during the policy life plus the remaining account
balance at maturity, regardless of the portfolio performance. Thus
even  if the account of the policyholder falls to zero before
maturity, GMWB feature will continue to provide the guaranteed
cashflows. GMWB allows the policyholder to withdraw funds below or
at contractual rate without penalty and above the contractual rate
with some penalty. If the policyholder behaves passively and the
withdrawal amount at each withdrawal date
 is predetermined at
the beginning of the contract, then the behavior of the
 policyholder is  called \emph{static}. In this case the paths of the account  can be simulated  and a standard Monte Carlo
 simulation method can be used to price the GMWB, though in low dimension problems partial differential equation (PDE) or integration methods are more efficient. On the other hand if the policyholder optimally decide the amount of withdrawal
 at each withdrawal date, then   the behavior of the policyholder is  called \emph{dynamic}.
A rational policyholder of GMWB will always choose the optimal
withdrawal strategy to maximize the present value of cashflows
generated from holding GMWB. Under the optimal withdrawal strategy
of a policyholder, the pricing of variable annuities with GMWB
becomes an optimal stochastic control problem. There is a rich
literature on dynamic programming to deal with optimal stochastic
control  problems in general, for textbook treatment see e.g. \cite{Powell2011, bauerle2011markov}. This
problem cannot be solved by a simulation based method such as the
well known Least-Squares Monte Carlo method introduced in \cite{Longstaff2001},
due to the fact that the paths of the underlying variable are altered by the
optimal withdrawal amounts at all pay dates prior to maturity and
thus they cannot be simulated. The variable annuities with
GMWB feature have been considered in e.g.
\cite{milevsky2006financial}, \cite{bauer2008universal},
\cite{dai2008guaranteed}, \cite{Forsyth2008},
\cite{bacinello2011unifying} and \cite{LuoShevchenko2014gmwb}.

In the case of a variable annuity contract with Guaranteed Minimum
Death Benefit,  the beneficiaries obtain a death benefit if the insured
dies during the deferment period. When variable annuities were
introduced, a very simple form of death benefit was predominant in
the market. Since the mid 1990s, insurers started to offer
a broad variety of death benefit designs.
The basic form of a death benefit is the so-called \emph{Return of Premium
Death Benefit}. Here, the maximum of the current account value at
time of death and the single premium is paid. Generally speaking,
given a mortality model or a Life Table,   the evaluation of GMDB
 is straightforward - a standard Monte Carlo method will work fine for such a problem and often closed form solutions can be obtained.

The pricing of GMWB is more involved, and it is even more
challenging under the dynamic (optimal) policyholder behavior.
  \cite{milevsky2006financial} developed a variety of methods for
  pricing GMWB products. In their {\it static} approach the GMWB product
  is decomposed into a Quanto Asian put plus a generic term-certain
  annuity. In their {\it dynamic} approach they assume the
  policyholder can terminate (surrender) the contract at the optimal time, which
  leads to an optimal stopping problem akin to pricing an American
  put option. \cite{bauer2008universal}
 presents valuation framework of variable annuities
with multiple guarantees. In their {\it dynamic} approach a strategy
  consists of  a numerous possible withdrawal amounts at each payment date.
They have developed a multidimensional discretization approach in
 which  the Black-Scholes PDE is  transformed to a
one-dimensional heat equation and a quasi-analytic solution is
obtained through a simple piecewise summation  with a  linear
interpolation on a mesh. Unfortunately the numerical formulation
considered in \cite{bauer2008universal} has four dimensions and the
computation of even a single  price of the GMWB contract under the
optimal policyholder strategy is very costly. It is mentioned in their paper that it took between 15 and 40 hours on the standard desktop PC to obtain a single price and no results for the dynamic case
were shown; also it looks like their methodology in the case of death benefit with dynamic GMWB corresponds to the upper estimator of the price (i.e. corresponds to formula (\ref{GMWDB_upper_eq}) in the next section). \cite{dai2008guaranteed} developed an efficient finite
difference algorithm using the penalty approximation to solve the
singular stochastic control problem for a continuous withdrawal
model under the dynamic withdrawal strategy.
 They have also developed a finite difference algorithm for the
  more realistic discrete withdrawal formulation.
  \cite{Forsyth2008}
present an impulse stochastic control formulation for pricing
variable annuities with GMWB under the optimal policyholder
behavior, and develop a single numerical scheme for solving the
Hamilton-Jacobi-Bellman variational inequality for the continuous
withdrawal model  as well as for pricing the discrete withdrawal
contracts.
 In \cite{bacinello2011unifying}
  the {\it static} valuations are performed via ordinary Monte Carlo method, and the  {\it mixed}
  valuations, where the policyholder is `semiactive' and can decide to  surrender the
 contract at any time during the life of the GMWB contract, are
 performed by the Least Squares Monte Carlo method.

 Recently we have developed a very efficient new  algorithm for pricing
 variable  annuities with GMWB under both static and dynamic policyholder behaviors solving an equivalent stochastic control problem; see \cite{LuoShevchenko2014gmwb}.
  Here the definition of
 ``dynamic" is similar to the one used by \cite{bauer2008universal},
   \cite{dai2008guaranteed} and \cite{Forsyth2008}, i.e. the rational policyholder can decide an
   optimal amount to withdraw at each payment date (based on information available at this date) to
   maximize the expected discounted value of the cashflows
   generated from holding the variable annuity with GMWB.
   The algorithm is neither based on  solving PDEs using finite difference nor on simulations
 using Monte Carlo. It  relies on computing the expected option values in a backward time-stepping between
 withdrawal dates through Gauss-Hermite integration quadrature applied on a cubic spline interpolation (referred to as GHQC). In \cite{LuoShevchenko2014gmwb} it is
  demonstrated that in pricing GMWB under the optimal policyholder behavior the GHQC algorithm can achieve similar accuracy
  as the finite difference method, but it is significantly faster because it requires less number of steps in time. This method can be applied when transition density of the underlying asset between withdrawal dates or it's moments are known in closed form and required expectations are 1d integrals. It has also been successfully used to price exotic options such as American, Asian, barrier, etc; see \cite{LuoShevchenkoGHQC2014}.

So far in the literature GMWB and GMDB are mainly considered  as two
separate contracts and it is implicitly assumed the policyholder
 of a GMWB contract will always live beyond the maturity date or there is always someone
there to make optimal withdrawal decisions for the entire duration
of the contract. In reality an elderly  policyholder may die before
maturity date, especially for a contract with a  long maturity. For
example, according to the Australian Life Table, Table \ref{tab_life}, a male aged 60 will
have more than $57\%$ probability to die before the age of 85. So,
for  a 60 year old male taking up a GMWB contract with a maturity of
25 years; the product design and pricing should certainly consider the
probability of death during the contract. Some variable annuity products may have expiry for the death benefit guarantee, e.g. at the age 70 or 75.

  In this paper we formulate pricing GMWDB (GMWB combined with GMDB) as a stochastic control problem, where at each withdrawal date the policyholder optimally decides the withdrawal amount based on information available at this date. It is important to note that pricing dynamic GMWDB for a given death time (i.e. conditional on knowing death time) and then averaging over possible death times according to death probabilities will lead to higher price than dynamic GMWDB where decisions are based on information available at withdrawal date (this will be discussed in the next section). We first extend the standard GMWB to allow for
 termination of the contract due to mortality risk,
returning the maximum of the remaining guarantee withdrawal amount
and the portfolio account value upon death.  In addition, two types
of extra death benefits are considered: paying the initial premium
or paying the maximum of the initial premium and the portfolio
account value upon death.
 Our recently developed
GHQC algorithm in \cite{LuoShevchenko2014gmwb}
  enables us to perform a comprehensive numerical study on the
  evaluation of GMWDB.

 In the next section  we describe the GMWDB product
  with extra death benefits
  on top of the usual GMWB features and outline the underlying stochastic model and corresponding
optimal stochastic control problem. Section \ref{sec_alg}
presents the numerical method and algorithm for pricing
GMWDB  under both static and dynamic policyholder behaviors.
 In Section \ref{NumericalResults_sec}
we present numerical results  for the fair fees
  of GMWDB  under a series of contract conditions. For the case of GMWDB
   under the optimal policyholder strategy, the inadequacy of the traditional fee structure
   is revealed in the case of additional death benefit. The fee cannot be charged as any proportion of the account value -- there is no solution to match the initial premium with the fair annuity price. An explanation of the inadequacy is given through an example
  which demonstrates why there is no solution for the fair fee for some long
maturities if the death benefit  paying at least the initial premium
is guaranteed to a rational policyholder. On the other hand, if a
fixed upfront fee or periodic instalment is charged, the extra fee
of
 adding life insurance to GMWB is cheaper than holding a separate
 life insurance. Concluding remarks are given in Section \ref{conclusion_sec}.

\section{Model and Stochastic Control Formulation}\label{sec_gmwb}

%
A variable annuity contract with
guaranteed minimum withdrawal benefit and death benefit (GMWDB) promises to return the entire initial investment through cash
withdrawals during the policy life plus the remaining account
balance at the contract maturity $T$, regardless of the portfolio performance. In addition, if policyholder dies before or at the contract maturity, then death benefit is paid to the beneficiaries.
We assume (common assumption in the academic research literature on pricing variables annuities) that market is complete in financial risk and mortality risk is completely diversified through selling enough policies and thus the annuity price can be expressed as expectation with respect to appropriate (risk-neutral) probability measure for the risky asset.
Below we outline the contract setup, model assumptions and solution via optimal stochastic control method.

\subsection{Assumptions and GMWDB contract details}
Assume that the annuity policyholder is allowed to take withdrawals $\gamma_n$ at times $t_1,\ldots, t_N=T$.
The premium paid upfront is invested into risky asset $S(t)$. Denote the value of corresponding \emph{wealth account} at time $t$ as $W(t)$, i.e. upfront premium is $W(0)$. GMWB is the guarantee of the
return of the premium via withdrawals $\gamma_n\ge 0$ allowed at
times $t_n$, $n=1,2,\ldots,N$.  Let $N_w$ denote the number of
withdrawals in a year (e.g. $N_w=12$ for a monthly withdrawal), then
the total number of withdrawals $N=\lceil\; N_w\times T \;\rceil$,
where $N=\lceil\; \cdot\;\rceil$ denotes  the ceiling of a float
number. The withdrawals cannot exceed the guarantee balance and
withdrawals can be different from contractual (guaranteed)
withdrawal amount $G_n=W(0)(t_n-t_{n-1})/T$ with penalties imposed if $\gamma_n>G_n$. Also denote the annual contractual rate as $g=1/T$.
Consider the following annuity contract details and model assumptions.

\begin{itemize}
\item \textbf{Risky asset process.} Let $S(t)$ denote the value of the reference portfolio of assets
(mutual fund index, etc.) underlying the variable annuity policy at
time $t$ that follows the risk neutral process
\begin{equation}\label{underlyingasset_process_eq}
dS(t)=r(t) S(t) dt+\sigma(t) S(t) dB(t),
\end{equation}
where $B(t)$ is the standard Wiener process, $r(t)$ is risk free
interest rate and $\sigma(t)$ is volatility.
Hereafter we assume that the model parameters are piecewise constant functions of time for time discretization
$0=t_0<t_1<\cdots<t_N=T$, where $t_0=0$ is today and $T$ is annuity contract maturity. Denote corresponding asset values as $S(t_0),\ldots,S(t_N)$ and risk free interest rate and volatility as $r_1,\ldots,r_N$ and $\sigma_1,\ldots,\sigma_N$ respectively. That is, $\sigma_1$ is the volatility for $(t_0,t_1]$; $\sigma_2$ is the volatility for $(t_1,t_2]$, etc. and similarly for interest rate.


\item \textbf{Wealth account.} For clarity, denote the value of the wealth account at time $t_n$ before withdrawal as $W(t_n^-)$ and after withdrawal as $W(t_n^+)$. Then, for the risky asset process (\ref{underlyingasset_process_eq}), the value of wealth account $W(t)$ evolves as
\begin{eqnarray}\label{wealth_process_eq}
W(t_n^-)&=&\frac{W(t_{n-1}^+)}{S(t_{n-1})}S(t_n) e^{-\alpha dt_n}=   W(t_{n-1}^+)e^{(r_n-\alpha-\frac{1}{2}\sigma^2_n)dt_n+\sigma_n \sqrt{dt_n} z_n},\\
W(t_n^+)&=&\max\left(W(t_n^-)-\gamma_n,0\right),\;\; n=1,2,\ldots,N,
\end{eqnarray}
where $dt_n=t_n-t_{n-1}$, $z_n$ are independent and identically distributed standard Normal random variables, and $\alpha$ is annual fee charged by insurance company. If the wealth account balance becomes zero or
negative, then it will stay zero till maturity. For the process (\ref{wealth_process_eq}), the transition density from $W(t_{n-1}^+)$ to $W(t_n^-)$ is lognormal density, i.e. $\ln W(t_{n-1}^+)$ is from Normal distribution with the mean $\ln W(t_n^-)+(r_n-\alpha-\frac{1}{2}\sigma^2_n)dt_n$ and variance $\sigma_n^2 dt_n$.

\item \textbf{Guarantee account.} Denote the value of \emph{guarantee account} at time $t$ as $A(t)$ with $A(0)=W(0)$, and the value of the account at $t_n$ before withdrawal as $A(t_{n}^{-})$ and after withdrawal as $A(t_{n}^{+})$. The guarantee account
evolves as
\begin{equation}\label{accountbalance_eq}
A(t_n^+)=A(t_{n}^{-})-\gamma_n=A(t_{n-1}^+)-\gamma_n,\;\; n=1,2,\ldots,N
\end{equation}
with $A(T^+)=0$ and $W(0)=A(0) \ge \gamma_1+\cdots+\gamma_N$ and
$A(t_{n-1}^+)\ge \sum_{k=n}^N\gamma_{k}$. Also note that guarantee account is unchanged within $(t_{n-1},t_n)$, i.e. $A(t_{n-1}^+)=A(t_{n}^-)$.

Some real products include ``step-up" arrangement that will increase guarantee account balance $A(t_n^-)$ to $\max(A(t_n^-),W(t_n^-))$ on anniversary dates, i.e. in the case of good investment performance. For simplicity we do not consider this feature explicitly but it is not difficult to incorporate this into the algorithm presented in this paper.

\item \textbf{Penalty.} The cashflow received by the policyholder at $t_n, n\in \{1,\ldots,N\}$ if alive is given by
\begin{equation}
C_n(\gamma_n)=\left\{\begin{array}{ll}
                   \gamma_n, & \mbox{if}\; 0\le \gamma\le G_n, \\
                   G_n+(1-\beta)(\gamma_n-G_n), & \mbox{if}\; \gamma_n>G_n,
                 \end{array} \right.
\end{equation}
where $G_n$ is contractual  withdrawal amount. That is, penalty is applied if
withdrawal $\gamma_n$ exceeds contractual amount $G_n$, i.e. $\beta\in
[0,1]$ is the penalty applied to the portion of withdrawal above
$G_n$.

To discourage excessive withdrawals beyond $G_n$, some contracts may include reset provision on the guarantee level $A(t_n^+)=\min(A(t_n^-),W(t_n^-))-\gamma_n$ if $\gamma_n>G_n$. We do not consider this explicitly but it is not a problem to incorporate this feature in the pricing algorithm presented in this paper.

\item \textbf{Death process.} Denote the time of policyholder death, a random variable, as $\tau$
with conditional death probabilities $q_n=\Pr[t_{n-1}<\tau \leq t_{n}|\tau>t_{n-1}]$. It is assumed that death time $\tau$ and asset process $S(t)$ are independent.
The policyholder age at $t_0$ is needed to find these probabilities from Life Tables. We assume that these probabilities are known and to simplify notation we do not use policyholder age variable explicitly in the formulas. In our numerical examples we assume age 60 for male and female policyholders and use the current Australian Life Table, see Table \ref{tab_life}.  Consider the corresponding Markov process defined by discrete random variables at $t_1,t_2,\ldots$
\begin{equation}\label{death_process_eq}
I_n=\left\{\begin{array}{cc}
               1, & \mbox{if policyholder is alive at $t_n$}, \\
               0, & \mbox{if policyholder died  during}\;(t_{n-1},t_n],\\
               -1, & \mbox{if policyholder died before, i.e.}\; \tau \le t_{n-1},
             \end{array}
  \right.
\end{equation}
with transition density from $I_{n-1}$ at $t_{n-1}$ to $I_n$ at $t_n$ specified by probabilities
$$
\hspace{-0.5cm}\begin{array}{lll}
\Pr[I_n=1|I_{n-1}=1]=1-q_n; &\Pr[I_n=1|I_{n-1}=0]=0; &\Pr[I_n=1|I_{n-1}=-1]=0;\\
\Pr[I_n=0|I_{n-1}=1]=q_n; &\Pr[I_n=0|I_{n-1}=0]=0; &\Pr[I_n=0|I_{n-1}=-1]=0;\\
\Pr[I_n=-1|I_{n-1}=1]=0; &\Pr[I_n=-1|I_{n-1}=0]=1; &\Pr[I_n=-1|I_{n-1}=-1]=1.
\end{array}
$$
For example, if death time is between $t_3$ and $t_4$, then realization of the process for $n=0,1,2,\ldots$ is $\bm{I}=\{1,1,1,1,0,-1,-1,\ldots\}$. Note that this variable $I_n$ is not affected by withdrawal at $t_n$.

\item \textbf{Death benefit.} If death time $\tau$ is after contract maturity $T$, then at the maturity the policyholder takes the maximum between the remaining guarantee withdrawal net of penalty charge and the remaining balance of the personal account,
i.e. the final payoff is
\begin{equation}\label{final_payoff_eq}
P_T(W(T^-),A(T^-))=\max\left(C_N\left(A(T^-)\right), W(T^-)\right).
\end{equation}

If death time $\tau$ occurs before or at contract maturity $T$ then the payoff taken by the beneficiary at death time slice $t_d$ (the first time slice larger or equal $\tau$) is the death benefit $P_D(W(t_d^-),A(t_{d}^-))$. We consider three types of death benefit denoted as DB0, DB1 and DB2 as follows:
\begin{equation}\label{deathbenefit_types}
P_D(W(t_d^-),A(t_{d}^-))=\left\{\begin{array}{cc}
                        \max\left(A(t_{d}^-), W(t_d^-)\right), & \mbox{death benefit DB0}, \\
                        W(0), & \mbox{death benefit DB1},\\
                         \max\left(W(0), W(t_d^-)\right), & \mbox{death benefit DB2}.
                      \end{array} \right.
\end{equation}
In the above, initial premium $W(0)$ is sometimes adjusted for inflation which is a trivial extension.

In some policies the death benefit may change at some age, for example DB2 or DB1 may change to DB0 at the age of 75 years (effectively making the death benefit guarantee expiring at the specified age). If death benefit expiring corresponds to switching to the standard GMWB, then it can be handled by setting death probabilities $q_n$ to zero after death benefit expiry till the contract maturity.

\item \textbf{Payoff.} Given withdrawal strategy $\bm{\gamma}=(\gamma_1,\ldots,\gamma_{N})$,
the present value of the overall payoff of the annuity contract is a function of state variables corresponding to wealth account $\bm{W}=(W(t_0),\ldots,W(t_N))$, guarantee account $\bm{A}=(A(t_0),\ldots,A(t_N))$ and death state $\bm{I}=(I_0,\ldots,I_N)$. Denote the state vector at time $t_n$ before the withdrawal as $X_n=(W(t_n^-),A(t_n^-),I_n)$ and $\bm{X}=(X_1,\ldots,X_N)$. Then the annuity payoff is

\begin{equation}\label{total_payoff_eq}
H_0(\bm{X},\bm{\gamma})= B_{0,N} h_N(X_N)+\sum_{n=1}^{N-1} B_{0,n} f_n(X_n,\gamma_n),
\end{equation}
where
\begin{equation}
h_N(X_N)=P_T(W(T^-),A(T^-))1_{\{I_N=1\}}+P_D(W(T^-),A(T^-))1_{\{I_N=0\}}
\end{equation}
is the cashflow at the contract maturity and
\begin{equation}
f_n(X_n,\gamma_n)= C_n(\gamma_n)1_{\{I_n=1\}}+ P_D(W(t_n^-),A(t_n^-))1_{\{I_n=0\}}
\end{equation}
is the cashflow at time $t_n$. Here, $1_{\{\cdot\}}$ is indicator function equals 1 if condition in $\{\cdot\}$ is true and zero otherwise, and $B_{i,j}$ is discounting factor from $t_j$ to $t_i$ \begin{equation}
B_{i,j}=\exp\left(-\int_{t_i}^{t_j}r(t)dt\right)=\exp\left(\sum_{n=i+1}^j r_n dt_n \right),\;t_j>t_i.
\end{equation}

\item \textbf{GMWDB static case.} Given the above assumptions and definitions the annuity price under the given \emph{static} strategy $\gamma_1,\gamma_2,\ldots,\gamma_{N}$ can be calculated as
\begin{equation}\label{GMWBstatic_eq}
Q_0\left(W(0),A(0),I_0=1\right)=\mathrm{E}_{t_0}^{\bm{X}}\left[H_0(\bm{X},\bm{\gamma})\right],
\end{equation}
where $\mathrm{E}_{t_0}^{\bm{X}}[\cdot]$ denotes expectation with respect to process $\bm{X}$ conditional on information available at time $t_0$. In the case of static strategy, $\bm{A}$ is deterministic and thus expectation is with respect to $\bm{W}$ and $\bm{I}$ processes.

\item \textbf{GMWDB dynamic case.} Under the optimal \emph{dynamic} strategy the annuity price is
\begin{equation}\label{GMWDB_general_eq}
Q_0\left(W(0),A(0),I_0=1\right)=\sup_{\bm{\gamma}}\mathrm{E}_{t_0}^{\bm{X}}\left[H_0(\bm{X},\bm{\gamma})\right],
\end{equation}
where $\gamma_n=\gamma_n(X_n)$ is a function of state variable $X_n$ at time $t_n$, i.e. can be different for different realizations of $X_n$. Note that in this case, $\bm{A}$ process is stochastic via stochasticity in $\bm\gamma$.

\item \textbf{GMWB case.}
 The standard GMWB contract corresponds to the above payoff for GMWDB if death probabilities are set to zero,  $q_1=\cdots =q_N=0$, i.e. $h_N(X_N)$ and $f_n(X_n,\gamma_n)$ in GMWDB payoff $H_0(\bm{X},\bm{\gamma})$ given in (\ref{total_payoff_eq}) simplify to $h_N(X_N)=P_T(W(T^-),A(T^-))$ and $f_n(X_n,\gamma_n)=C_n(\gamma_n)$. Then static and dynamic GMWB prices are given by (\ref{GMWBstatic_eq}) and (\ref{GMWDB_general_eq}) respectively where expectations are calculated with respect to $\bm{W}$ process.

\end{itemize}

\subsection{Solving Optimal Stochastic Control Problem}
Given that state variable $\bm{X}=(X_1,\ldots,X_N)$ is Markov process, it is easy to recognize that the annuity valuation under the optimal withdrawal strategy (\ref{GMWDB_general_eq}) is optimal stochastic control problem  for Markov process that can be solved  recursively to find annuity value $Q_{t_n}(x)$ at $t_n$, $n=N-1,\ldots,0$ via backward induction
\begin{equation}
Q_{t_n}(x)=\sup_{0\le\gamma_n\le A(t_{n}^-)}\left(f_n(X_n,\gamma_n(X_n))+ B_{n,n+1}\int Q_{t_{n+1}}(x^\prime)K_{t_n}(dx^\prime|x,\gamma_n) \right)
\end{equation}
starting from final condition $Q_T(x)=h_N(x)$. Here $K_{t_n}(dx^\prime|x,\gamma_n)$ is  the stochastic kernel representing probability to reach state in $dx^\prime$ at time $t_{n+1}$ if the withdrawal (\emph{action}) $\gamma_n$ is applied in the state $x$ at time $t_n$. For a good textbook treatment of stochastic control problem in finance, see \cite{bauerle2011markov}.

Explicitly, this backward recursion can be rewritten as
\begin{eqnarray}
&&\hspace{-1cm}{Q}_{t_n^+}\left(W,A,I_n\right)=
B_{n,n+1}\mathrm{E}_{t_n}^{X_{n+1}}\left[{Q}_{t_{n+1}^{-}}\left(W(t_{n+1}^{-}),A(t_{n+1}^{-}),I_{n+1}\right)\left|
W,A,I_n\right.\right],\nonumber
\\\label{continuation_value_eq}
&&\hspace{-1cm}{Q}_{t_n^{-}}\left(W,A;I_n\right)=
\max_{0\le\gamma_{n}\le A}\left( C_n(\gamma_n)1_{\{I_n=1\}}+P_D(W,A)1_{\{I_n=0\}}+{Q}_{t_n^+}((W-\gamma_n,0),A-\gamma_n, I_n)\right)\nonumber
\label{jump_cond_eq}
\end{eqnarray}
for $n={N}-1,{N}-2,\ldots,0$ starting from maturity condition at $t_N=T$
\begin{equation}
{Q}_{t_{{N}}^{-}}\left(W(t_{{N}}^{-}),A(t_{{N}}^{-}),I_N\right)=P_T(W(T^-),A(T^-))1_{\{I_N=1\}}+
P_D(W(T^-),A(T^-))1_{\{I_N=0\}},
\end{equation}
where for clarity we denote ${Q}_{t_n^-}(\cdot)$ and ${Q}_{t_n^+}(\cdot)$ the annuity values at time $t_n$ before and after withdrawal respectively.
Taking expectation with respect to death variable $I_{n+1}$ and using ${Q}_{t_n^+}\left(W,A,I_n=0\right)={Q}_{t_n^+}\left(W,A,I_n=-1\right)={Q}_{t_n^{-}}\left(W,A;I_n=-1\right)=0$ and ${Q}_{t_{n}^{-}}\left(W(t_{n}^{-}),A(t_{n}^{-}),I_{n}=0\right)=P_D(W(t_n^-),A(t_n^-))$,  it simplifies to the recursion equations
\begin{eqnarray}\label{GMWDB_recursion_eq}
&&\hspace{-1.5cm}{Q}_{t_n^+}\left(W,A,I_n=1\right)\nonumber\\
&=&(1-q_n)B_{n,n+1}\mathrm{E}_{t_n}^{X_{n+1}}\left[{Q}_{t_{n+1}^{-}}\left(W(t_{n+1}^{-}),A(t_{n+1}^{-}),I_{n+1}=1\right)\left|
W,A,I_n=1\right.\right]\nonumber\\
&&+q_nB_{n,n+1}\mathrm{E}_{t_n}^{X_{n+1}}\left[P_D(W(t_{n+1}^{-}),A(t_{n+1}^{-}))\left|
W,A,I_n=1\right.\right]\nonumber\\
\end{eqnarray}
with jump condition
\begin{equation}
\label{GMWDB_jump_condition_eq}
{Q}_{t_n^{-}}\left(W,A;I_n=1\right)=
\max_{0\le\gamma_{n}\le A}\left( C_n(\gamma_n)+{Q}_{t_n^+}((W-\gamma_n,0),A-\gamma_n, I_n=1)\right),
\end{equation}
and maturity condition
\begin{equation}\label{GMWDB_final_condition_eq}
{Q}_{t_{{N}}^{-}}\left(W(t_{{N}}^{-}),A(t_{{N}}^{-}),I_N=1\right)=P_T(W(T^-),A(T^-)).
\end{equation}
Note that expectations in (\ref{GMWDB_recursion_eq}) are with respect to $W(t_{n+1}^-)$ only because $A(t_n^+)=A(t_{n+1}^-)$.

\subsection{Dynamic GMWDB Upper and Lower Estimators}
There are many lower estimators that can be constructed for dynamic GMWDB price $Q_{t_0}(\cdot)$ given by (\ref{GMWDB_general_eq}). In particular, any static (deterministic) strategy will produce lower estimator given by equation (\ref{GMWBstatic_eq}). In Section \ref{NumericalResults_sec} we will utilize one of such lower estimators.

One of the possible upper estimators of
the dynamic GMWDB (\ref{GMWDB_general_eq}) can be found via calculations of GMWDB conditional on death time (i.e. perfect forecast of the death time). For easier understanding and notational convenience, instead of dealing with the death process $I_n$ in the following we consider the corresponding death time random variable $\tau$ and let $t_d$   be the
first withdrawal time equal or exceeding $\tau$   and
$\widetilde{N}=\min(d,N)$ (if $\tau>T$ then without loss of
generality set $d=N+1$). To avoid confusion, we denote this GMWDB conditional on $\tau$ by $V_0(\cdot;t_d)$. Conditional on $\tau$ (i.e. conditional on $t_d$), it is given by
\begin{eqnarray}\label{eq_Vs}
V_{0}\left(W(0),A(0);t_d\right)=\max_{\gamma_{1},\ldots,\gamma_{\widetilde{N}-1}}\mathrm{E}_{t_0}^{\bm{W}}
\left[
B_{0,\widetilde{N}} \widetilde{h}(t_{\widetilde{N}})+\sum_{n=1}^{\widetilde{N}-1} B_{0,n} C_n(\gamma_n)
\right],
\end{eqnarray}
$$
\widetilde{h}(t_{\widetilde{N}})=P_T(W(t_N^-),A(t_N^-))1_{\{\tau>T\}}+
P_D(W(t_d^-),A(t_d^-))1_{\{\tau\le T\}},
$$
where $\mathrm{E}_{t_0}^{\bm{W}}[\cdot]$ is expectation with respect to wealth process $W(t_n),n=0,1,\ldots,N$. It can be solved
using backward recursion
\begin{eqnarray}
&&\hspace{-1cm}{V}_{t_n^+}\left(W,A;t_d\right)=
e^{-r_n(t_{n+1}-t_n)}\mathrm{E}_{t_n}^{W(t_{n+1}^-)}\left[{V}_{t_{n+1}^{-}}\left(W(t_{n+1}^{-}),A(t_{n+1}^{-});t_d\right)\left|
W,A;t_d\right.\right],\nonumber
\\\label{continuation_value_upper_estimator_eq}
&&\hspace{-1cm}{V}_{t_n^{-}}\left(W,A;t_d\right)=
\max_{0\le\gamma_{n}\le A}\left( C_n(\gamma_n)+{V}_{t_n^+}((W-\gamma_n,0),A-\gamma_n;t_d )\right)\nonumber
\label{jump_cond_upper_estimator_eq}
\end{eqnarray}
for $n=\widetilde{N}-1,\widetilde{N}-2,\ldots,0$ starting from maturity condition at $\widetilde{N}=\min(d,N)$
\begin{equation}
{V}_{t_{\widetilde{N}}^{-}}\left(W(t_{\widetilde{N}}^{-}),A(t_{\widetilde{N}}^{-});t_d\right)=\widetilde{h}(t_{\widetilde{N}}).
\end{equation}
Then the expectation with respect to death time gives
\begin{eqnarray}\label{GMWDB_upper_eq}
{Q}_{0}^{(\mathrm{u})}\left(W(0),A(0),I_0=1\right)&=&\mathrm{E}_{t_0}^{\tau}[{V}_{0}\left(W(0),A(0);t_d\right)]\nonumber\\
&=&\Pr[\tau>T]{V}_{0}\left(W(0),A(0);t_{N+1}\right)\nonumber\\
&&+\sum_{n=1}^N \Pr[t_{n-1}<\tau\le t_n]{V}_{0}\left(W(0),A(0);t_n\right).
\end{eqnarray}
Note that ${V}_{0}\left(W(0),A(0);t_{N+1}\right)$ corresponds to the standard GMWB (i.e. GMWDB conditional on death after contract maturity). Also note that the death probabilities $p_n=\Pr[t_{n-1}<\tau\le t_n]$ are different from conditional death probabilities $q_n=\Pr[t_{n-1}<\tau\le t_n|\tau>t_{n-1}]$ in the death process (\ref{death_process_eq}); both $p_n$ and $q_n$ are easily found from Life Tables or mortality process models.
${Q}_{0}^{(\mathrm{u})}(\cdot)$ is the upper estimator for dynamic GMWDB $Q_{t_0}(\cdot)$, given by (\ref{GMWDB_general_eq}),
because it is based on optimal strategy conditional on perfect forecast for time of death for each death process realization. Formally,

\begin{eqnarray}\label{GMWDB_upper_inequality}
&&Q_0^{(\mathrm{u})}\left(W(0),A(0),I_0=1\right)=\mathrm{E}_{t_0}^{\tau}\left[\sup_{\bm{\gamma}}\mathrm{E}_{t_0}^{\bm{W}}\left[\left. H_0(\bm{X},\bm{\gamma(W,A)})\right|\tau \right]\right]\nonumber\\
&&\hspace{1cm}\ge \sup_{\bm{\gamma}}\mathrm{E}_{t_0}^{\tau,\bm{W}}\left[H_0(\bm{X},\bm{\gamma}(\bm{X}))\right]=Q_0\left(W(0),A(0),I_0=1\right).
\end{eqnarray}

\section{Numerical algorithm}\label{sec_alg}
In the literature, the optimal stochastic control problem for pricing
GMWB with discrete optimal withdrawals has only been successfully
dealt with by solving the one-dimensional PDE equation using a finite difference method presented in \cite{dai2008guaranteed} and \cite{Forsyth2008}. Simulation based method such as the Least Squares Monte Carlo method cannot be
  applied for such problems due to the dynamic behavior of the policyholder affecting
  the paths of the underlying wealth account.
 Recently, \cite{LuoShevchenko2014gmwb}  have considered an alternative method
without solving  PDEs or simulating paths.
 The new approach  relies on computing the expected option values in a backward time-stepping between
 withdrawal dates through a high order Gauss-Hermite integration quadrature applied on a cubic spline
 interpolation. In this paper we adopt this method to calculate GMWDB.


\subsection{Numerical quadrature to evaluate the expectation}\label{GHQCalgo_sec}
To price variable annuity contract
with GMWDB, i.e. to compute $Q_0\left(W(0),A(0),I_0=1\right)$, we have
to evaluate expectations in (\ref{GMWDB_recursion_eq}).
Assuming the conditional probability
 density of $W(t_n^-)$ given $W(t_{n-1}^+)$  is known as
$\tilde{p}_n(w|W(t_{n-1}^+))$, in the case of process (\ref{wealth_process_eq}) it is just lognormal density, (\ref{GMWDB_recursion_eq}) can be evaluated
as

\begin{equation}\label{eq_intS}
{Q}_{t_{n-1}^+}\left(W(t_{n-1}^+), A,I=1\right)=B_{n-1,n}\int_0^{+\infty}
\tilde{p}_n(w|W(t_{n-1}^+)) \widetilde{Q}_n(w) dw,
\end{equation}
where
$$
\widetilde{Q}_n(w)=B_{n-1,n}\left((1-q_n)Q_{t_n^-}(w,A,I=1)+q_n P_D(w,A)\right).
$$
We use Gauss-Hermite quadrature for the evaluation of the
above integration over an infinite domain. The required continuous
function $Q_t(w,\cdot)$ will be approximated by a cubic spline
interpolation on a discretized grid in the  $W$ space.

The wealth account domain is discretized as $W_{\min}
=W_0 < W_1, \ldots,W_M=W_{\max}$, where $W_{\min}$ and $W_{\max}$
are the lower and upper boundary, respectively. For pricing GMWDB, because of the finite reduction of $W$ at each
withdrawal date, we have to consider the possibility of $W$ goes to zero, thus the lower bound $W_{\min} = 0$. The upper bound is set sufficiently far from the spot asset value at
time zero W(0). A good choice of such a boundary could be $W_{\max} = W(0) \exp(5\sigma T)$.  The idea is to
find annuity values at all these grid points at  each time step
$t_n$  through integration  (\ref{eq_intS}),
starting at maturity $t_N^-=T^-$.  At each time step we evaluate
the integral (\ref{eq_intS}) for every grid point  by a high
accuracy numerical quadrature.

The annuity value at
$t=t_n^-$ is known only at grid points $W_m$, $m=0,1,\ldots,M$. In
order to approximate the continuous function $Q_t(w,\cdot)$ from the
values at the discrete grid points, we
 use the natual cubic spline interpolation
 which is smooth in the first derivative and continuous in the second derivative; and second derivative is zero for extrapolation region.

The process for $W(t)$  between withdrawal dates is a simple
 process given in (\ref{wealth_process_eq}), the conditional density of
$W(t_n^-)$ given $W(t_{n-1}^+)$  is from a lognormal distribution.
To apply Gauss-Hermite numerical
quadrature for integration over an infinite domain,
we introduce a
new variable
   \begin{equation}\label{eq_y}
Y(t_n)=\frac{\ln\left(W(t_n^-)/W(t_{n-1}^+)\right)-(r_n-\alpha-\frac{1}{2}\sigma^2_n )dt_n}{\sigma_n
\sqrt{dt_n}},
 \end{equation}
and denote the function $\widetilde{Q}_n(w)$ after this transformation as $\widetilde{Q}_n^{(y)}(y)$. Then the integration becomes
  \begin{equation}\label{eq_qy}
{Q}_{ t_{n-1}^+}\left(W(t_{n-1}^+), A,I_n=1\right)=\frac{
1}{\sqrt{2\pi}} \int_{-\infty}^{+\infty} e^{-\frac{1}{2}y^2}\widetilde{Q}^{(y)}_n(y)dy.
\end{equation}
For an arbitrary function $f(x)$, the Gauss-Hermite
quadrature is
\begin{equation}\label{eq_GHQ}
\int_{-\infty}^{+\infty}e^{-x^2}f(x)dx \approx \sum_{j=1}^q
\lambda_j^{(q)} f(\xi_j^{(q)}),
\end{equation}
where  $q$ is the order of the Hermite polynomial, $\xi_i^{(q)}, i = 1,2,\ldots,q$ are
the roots of the Hermite polynomial $H_q(x)$, and
the associated weights $ \lambda_j^{(n)}$  are given by

$$\lambda_i^{(q)}= \frac {2^{q-1} q! \sqrt{\pi}} {q^2[H_{q-1}(\xi_i^{(q)})]^2}.$$

Applying a change of variable $x=y/\sqrt{2}$ and using the
Gauss-Hermite quadrature to (\ref{eq_qy}), we obtain

 \begin{equation}\label{eq_qX}
{Q}_{t_{n-1}^+}\left(W(t_{n-1}^+), A,I_n=1\right) \approx  \frac{
1}{\sqrt{\pi}} \sum_{i=1}^q \lambda_i^{(q)}
\widetilde{Q}_n^{(y)}(\sqrt{2}\sigma_n
\sqrt{dt_n} \xi_j^{(n)}).
\end{equation}

If we apply the change of variable (\ref{eq_y}) and the
Gauss-Hermite quadrature (\ref{eq_qy}) to every grid point $W_m$,
$m=0,1,\ldots,M$, i.e. let $W(t_{n-1}^+)=W_m$, then the option
values at time $t=t_{n-1}^+$  for all the grid points  can be
evaluated through (\ref{eq_qX}).
If the transition density function from $W(t_{n-1}^+)$ to $W(t_n^-)$ is not known in closed form but one can find its moments, then integration can also be  done by matching moments as described in \citet{LuoShevchenkoGHQC2014,LuoShevchenko2014gmwb}.

\subsection{Jump condition application}
Any change of $A(t)$ only occurs at withdrawal dates. After the
amount $\gamma_n$ is drawn at $t_n$, the annuity account reduces
from $W(t_n^-)$ to $W(t_n^+) = \max (W(t_n^-) -\gamma_n,0)$, and the
guarantee balance drops from $A(t_n^-)$ to $A(t_n^+)=A(t_n^-) -
\gamma_n$.
 The jump condition of $Q_t(W,A,I_n=1)$ across $t_n$ is given
by

\begin{equation}\label{eqn_jump}
Q_{t_n^-}(W,A,I_n=1)=\max_{0 \leq \gamma_n\leq A } [Q_{t_n^+}(\max(W-\gamma_n,0),
A-\gamma_n,I_n=1)+C(\gamma_n)].
\end{equation}
For the optimal strategy, we chose a value for $\gamma_n$ under the
restriction $0 \leq \gamma_n\leq A $ to maximize the function value
$Q_{t_n^-}(W,A,I_n=1)$ in (\ref{eqn_jump}).

To apply the jump conditions,  an auxiliary finite grid $0 = A_1 <
A_2 < \cdots < A_J = W(0)$ is introduced to track the remaining
guarantee balance $A$, where $J$ is the total number of nodes in the
guarantee balance amount coordinate. For each $A_j $, we associate a
continuous solution from  (\ref{eq_qX}) and the cubic spline
interpolation.
  We can limit the number of
possible discrete withdrawal amounts to be finite by only allowing
the guarantee balance to be equal to one of the grid points $0 = A_1
< A_2 < \cdots < A_J = W(0)$. This implies that, for a given balance
$A_j$ at time $t_n^-$,  the withdraw amount $\gamma$ takes $j$
possible values: $\gamma=A_j-A_k$, $k=1,2,\ldots,j$ and  the jump
condition (\ref{eqn_jump}) takes the following form

\begin{equation}\label{eqn_jump2}
Q_{t_n^-}(W_m,A_j,I_n=1)=\max_{1\leq k \leq j} [Q_{t_n^+}(\max(W_m-A_j+A_k,0), A_k,I_n=1)+C(A_j-A_k)].
\end{equation}

For the optimal strategy, we chose a value for  $1 \leq  k \leq j $ to
maximize  $Q_{t_n^-}(W_m,A_j,I_n=1)$ in (\ref{eqn_jump2}).
Note that although the jump amount $\gamma=A_j-A_k$,
$k=1,2,\ldots,j\;$ is independent of time $t_n$ and account value
$W_m$, the value $Q_{t_n^+}(\max(W_m-A_j+A_k,0), A_k,I_n=1)$ depends on all
variables $(W_m,A_j,t_n)$ and the jump amount.

It is worth pointing out that part of the good efficiency of the
GHQC algorithm for pricing GMWB or GMWDB under rational policyholder behavior
is due to that fact that the same cubic spline interpolation is used
for both numerical integration (\ref{eq_intS}) and the application
of jump condition (\ref{eqn_jump2}). A clear advantage of this
 numerical algorithm over PDE based finite difference
approach is that significantly smaller number of time steps are
required because the transition density
over the finite time period in  (\ref{eq_intS}) is known.
The finite difference method requires dividing the
period between two consecutive withdrawal dates into finer time
steps for a good accuracy due to the finite difference approximation
to the partial derivatives.


\subsection{Death probabilities}
Given a Life Table, such as Table \ref{tab_life}, estimating the death probabilities
$q_n$ and $p_n$ required in (\ref{death_process_eq}) and (\ref{GMWDB_upper_eq})
is straightforward.
The Life Table tabulates the number of people surviving to the exact age starting with 100,000 at the age zero for each sex and goes beyond 100 years. Denote the number of people still alive at age
$k$ years as a list $L(k)$, $k=0,1,\ldots,K$. Denote the age of policyholder at the start of the contract (i.e. at time $t_0=0$) as $k_0$. To
estimate the conditional death probabilities $q_n=\Pr[t_{n-1}<\tau
\leq t_{n}|\tau>t_{n-1}]$ and $p_n=\Pr[t_{n-1}<\tau \leq
t_{n}|\tau>t_{0}]$, we only need to know the number of people alive
at time $t=t_{n-1}$ and $t=t_n$. Because in the Life Table the
number of people $L(k)$ is only given at integer number $k$, we estimate the number of people alive $L(k_0+t)$ for an arbitrary time $t$ using linear interpolation, i.e. assuming a uniform distribution
for the death time within a year. Of course a more elaborate
interpolation is also possible. Then, for $k\leq t+k_0 \leq k+1$, $L(k_0+t)$ is calculated as
$$L(k_0+t)=(k+1-t-k_0)\times L(k)+(t+k_0-k)\times L(k+1).$$
\noindent Having obtained $L(k_0+t_{n-1})$ and $L(k_0+t_n)$ using the above
formula (note $t_{n-1}$ and $t_n$ may not necessarily fall within
the same year), the conditional death probabilities are estimated as
\begin{equation}
\begin{split}
&q_n=\Pr[t_{n-1}<\tau \leq t_{n}|\tau>t_{n-1}] \approx \frac{L(k_0+t_{n-1})-
L(k_0+t_n)}{L(k_0+t_{n-1})},\\
&p_n=\Pr[t_{n-1}<\tau \leq t_{n}|\tau>t_{0}] \approx \frac{L(k_0+t_{n-1})-
L(k_0+t_n)}{L(k_0+t_0)}.
\end{split}
\end{equation}
 Instead of a Life Table, stochastic
mortality models such as frequently used benchmark Lee-Carter model (\cite{Lee-Carter}) forecasting mortality rate using stochastic process (and typically accounting for systematic mortality risk) can also be used for estimating death probabilities $q_n$ and $p_n$.

\subsection{Overall algorithm description}

Starting from a final condition
at $t=T^-$ (just immediately before the final withdrawal), a
backward time stepping using (\ref{eq_intS}) gives solution up to
time $t=t_{N-1}^+$. Applying jump condition (\ref{eqn_jump}) to the
solution at $t=t_{N-1}^+$ we obtain the solution at $t=t_{N-1}^-$
from which further backward time stepping gives us solution at
$t=t_{N-2}^+$, etc till $t_0$. The numerical algorithm takes the
following key steps

~

\begin{algorithm_new}[GMWDB pricing]\label{algorithm1}

 ~

\begin{itemize}
\item Step 1. Generate an auxiliary finite grid  $0 = A_1 < A_2 <
\cdots < A_J = W(0)$ to track the guarantee  account $A$.
\item Step 2.  Discretize wealth account $W$ space as $W_0
,W_1, \ldots,W_M$ which is a grid for computing (\ref{eq_intS}).
\item Step 3. At $t=t_N=T$ apply the final condition at each node point $(W_m, A_j)$, $j=1,2,\ldots, J$, $m=1,2,\ldots, M$
to get payoff $Q_{T^-}(W,A,I=1)$.
\item Step 4. Evaluate integration (\ref{eq_intS}) (for each of the $A_j$)  to obtain $Q_{t_{N-1}^+}(W, A,I_n=1)$.
\item Step 5. Apply the jump condition (\ref{eqn_jump}) to obtain $Q_{t_{N-1}^-}(W, A,I_n=1)$ for all possible jumps $\gamma$ and find the jump that maximizes
$Q_{t_{N-1}^-}(W, A,I_n=1)$.
\item Step 6. Repeat Step 4 and 5 for $t=t_{N-2}, t_{N-3}, \ldots, t_1$.
\item Step 7. Evaluate integration (\ref{eq_intS}) for the backward time step from $t_1$ to $t_0$ for the single node value
$A=A_J=W(0)$ to obtain solution  $Q_0(W, A_J,I_0=1)$ and take the value
$Q_{t_0}(A_J, A_J, I_0=1)$ as the annuity price at $t=t_0$. Of course if the contract is re-evaluated at some time after it started, one will need to take solution at the node corresponding to $A$ and $W$ at that time.
\end{itemize}
\end{algorithm_new}

We use Gauss-Hermite quadrature integration (\ref{eq_qX}) with cubic spline in Step 4. One can also perform integration using moment matching if the density is not known in closed form but its moments are available.
 For static case, Step 1
is not needed because only a single solution is required and the
jump condition applies to the single solution itself.

\section{Numerical  Results}\label{NumericalResults_sec}
The accuracy and efficiency  of GHQC method in pricing GMWB under optimal
withdraw is well demonstrated in \cite{LuoShevchenkoGHQC2014}. From
numerical point of view, once the problem is correctly formulated,
pricing GMWDB (with combined GMWB and GMDB features) uses the same
key algorithm components as those for pricing GMWB, such as
numerical quadrature integration for the expectations and cubic
spline interpolation for applying the jump conditions.

To the best of our knowledge, there is no numerical results
available in the literature for
  variable annuity contracts with combined GMWB and GMDB features under the optimal withdrawal strategy, and only some very limited
results are available in the literature for GMWB under the optimal
withdrawal strategy, namely from \cite{dai2008guaranteed} and
\cite{Forsyth2008}. To
validate our implementation for GMWDB, we have made the
  following efforts:
  \begin{itemize}
  \item Implemented an efficient  finite difference (FD) algorithm with the same
  mathematical formulation and functionality, so GHQC results for GMWDB can always be compared with FD.
  \item For the limiting case where the death probability is zero,
  i.e. $q_n=0,\;n=1,\ldots,N$, the GMQDB price should reduce to the standard
  GMWB price exactly (still under optimal withdrawal), and we can compare
  our GMWDB results with GMWB results found in the literature.
  \item For the static withdraw case, GMWDB contract can be evaluated by
  Monte Carlo (MC) method, so we can compare our GHQC results for GMWDB with
   those of MC using a very large number of simulations. The close agreement of results between the two algorithms offers a reassuring validation for both methods.

  \end{itemize}

Below we present and discuss numerical results for GMWDB.
In our numerical examples we assume policyholder age 60 for male and female and use the current Australian Life Table, see Table \ref{tab_life}, to find corresponding death probabilities.

\subsection{Results for dynamic GMWB}\label{sec_result1}
 Before showing numerical results for GMWDB, we first
validate our numerical algorithms by testing the limiting case of
zero death probability. In such a case we can compare our GMWDB
results with some finite difference results found in the literature
for GMWB. In \cite{Forsyth2008}, the fair fees for the
  discrete withdrawal model with $g=10\%$ for the yearly ($N_w=1$) and
  half-yearly ($N_w=2$) withdrawal frequency at $\sigma=0.2$ and $\sigma=0.3$  were presented in a carefully performed
  convergence study, with the same values for other input
  parameters ($r=5\%$, $\beta=10\%$). Table \ref{tab_g10} compares
  GHQC ({Algorithm \ref{algorithm1}}) results with those of \cite{Forsyth2008}. The values of
  \cite{Forsyth2008} quoted in Table \ref{tab_g10} correspond to their finest mesh grids and time
  steps at $M=2049$ grid points for $W$ coordinate, $J=1601$ grid points for $A$ coordinate and $1920$ steps for time, while  our GHQC values were
  obtained using $M=400$, $J=100$ and with $q=9$ for the number of quadrature
  points (and the number of time steps is the same as the number of withdrawal dates $N$). As shown in Table \ref{tab_g10}, the maximum absolute
  difference in the fair fee rate between the two numerical studies is only $0.3$ basis point, and the average absolute
   difference of the four cases in the table is less than 0.2 basis
   point. As one basis point is $0.01\%$, the average difference is in the order of 20 cents for a one thousand dollar account.
    In relative terms, the maximum difference is less than $0.15\%$, and the average magnitude of the relative differences between
   the two studies is less than $0.08\%$. \cite{Forsyth2008} did not provide CPU numbers for
   their
   calculations of fair fees. In our case each calculation of the
   fair fee in Table \ref{tab_g10}, which involves a Newton iterative
   method of root finding, took about 5 seconds. In \cite{LuoShevchenko2014gmwb}, a
   detailed comparison shows the GHQC algorithm is significantly
   faster than the FD in pricing dynamic GMWB contract, especially for lower withdrawal frequencies.
   All our calculations in this study were performed using standard desktop PC (Intel(R)  Core(TM) i5-2400 @3.1GHz).

\begin{table}[!h]
\begin{center}
{\footnotesize{\begin{tabular*}{0.85\textwidth}{cccc} \toprule
 \bf{withdrawal frequency} & \bf{volatility, }$\sigma$ &  \bf{Fair fee, $\alpha$} &  \bf{Fair fee}, $\alpha$  \\
  &                       &\cite{Forsyth2008} & \bf{GHQC} \\
 \midrule
 yearly      & 0.2 & 129.1 & 129.1   \\
 half-yearly & 0.2 & 133.5 & 133.7 \\
 yearly      & 0.3 & 293.3 & 293.5  \\
 half-yearly & 0.3 & 302.4 & 302.7   \\
\bottomrule
\end{tabular*}
}}\end{center} \caption{\footnotesize\bf Fair fee $\alpha$ in bp ($1\mathrm{bp}=0.01\%$) for GMWB annuity with optimal policyholder withdrawals (i.e. equivalent to GMWDB with death probability set to zero). Comparison of results obtained by GHQC method and those from finite difference by \cite{Forsyth2008}. The input parameters are $g=10\%$, $\beta=10\%$ and
$r=5\%$.} \label{tab_g10}
\end{table}

\subsection{Results for GMWDB}\label{sec_result2}
The standard GMWB annuity without consideration of death event assumes that the policyholder will be alive during the contract or there is beneficiary to continue withdrawals until maturity. We refer to this case as \emph{GMWDB without death benefit}; it is equivalent to GMWDB with death probability set to zero.
There are several more or less natural considerations of GMWDB contact payoff in the case of death before or at contract maturity. We consider death benefit types summarised in (\ref{deathbenefit_types}), though many other modifications are possible. A natural contract condition in the case of death event is to pay the maximum of the remaining
guaranteed account and the wealth account values (denoted as DB0). An alternative is to payback
the initial premium at death (denoted as DB1), which is similar to a term life
insurance paying a fixed amount equal to the premium upon death (the amount can be adjust for inflation, etc). On the other hand, the Guaranteed Minimum Death Benefit (GMDB), in its
basic form, offers to pay the maximum of the wealth account value and the
premium (denoted as DB2).
GMWDB contract with DB2 has both features of GMWB and GMDB but requiring
only a single premium. All three types of death benefits can be
added to either static GMWB or dynamic GMWB. All the following
results of GMWDB are based on death probabilities calculated from the Life Table for the male population
in Australia, except those in Table \ref{tab_female}  which have
used the Life Table for the female population in Australia; also see Table \ref{tab_life}. We assume that policyholder is 60 years old.

\subsubsection{Static GMWDB}
Table \ref{tab_feeSD1} shows static case results of GMWDB with DB0, DB1 and DB2 death
benefits, and standard GMWB (i.e. GMWDB with zero death probability). Here we also present MC and FD results for DB0 case. The number of simulations for MC
  is 20 million and corresponding standard error in calculation of fair fees is within 0.2 basis point. As shown in Table \ref{tab_feeSD1}, the maximum difference
 of fair fee between GHQC and MC methods is 0.2 basis point, and the maximum difference
 between GHQC and FD methods is 0.1 basis point in the case of DB0. We have also calculated DB1 and DB2 cases using FD and MC methods and the difference between the methods is virtually the same as in the case of DB0 so we do not present these results. The close agreement between
 MC  and  GHQC methods is a reassuring
 validation for both algorithms.

Comparing with results for static GMWB (i.e. GMWDB with death probability set to zero), adding DB0 death benefit requires less than 10 extra basis points in the fair fee.
The fair fee increases significantly in the case of DB2 death benefit. As expected, the fair fee for
DB2 at all maturities is  higher than for DB1 and the difference decreases as maturity increases. At
$g=4\%$ ($T=25$ years), the fair fee for DB1 is actually negative,
meaning the contract is not appropriate. This
seems to be odd at first, but it actually makes sense: the $g=4\%$
withdrawal rate is lower than the expected growth rate of $r=5\%$, and
returning only the premium at any time of death is a loss to the
policyholder, and a possible gain can only be obtained if the
policyholder survives beyond the maturity which is not enough to
offset the loss due to the probability of death. At $g=r=5\%$, the
fair fee for DB1 becomes positive but it is still lower than for
DB0, and it is even lower than a static GMWB.

Real product design may impose expiry date for the death benefit guarantee  (e.g. at the age 70 or 75) that will reduce the cost of the death benefit guarantee itself and will help avoid the case of negative fee such as in Table \ref{tab_feeSD1} for long contract maturity $T=25$ years. If the contract will switch from DB2 or DB1 to DB0 (effectively making the death benefit expiring at the specified age), then it can be handled by the same algorithm described in this paper with adjustment to the death benefit function (\ref{deathbenefit_types}). If the death benefit expiring corresponds to switching to the standard GMWB, then it can be handled by setting death probabilities $q_n$ to zero after death benefit expiry till the contract maturity.

\begin{table}[!h]
\begin{center}
{\footnotesize{\begin{tabular*}{0.95\textwidth}{cccccccc} \toprule
\bf{contractual rate} & \bf{maturity} & \bf{GHQC}  &  \bf{GHQC} &   \bf{FD}  & \bf{MC} & \bf{GHQC} & \bf{GHQC} \\
$g$ & $T=1/g$ & no death & DB0  & DB0 & DB0 & DB1 & DB2\\
 \midrule
 $4\%$ & 25& 17.69 & 25.53 & 25.49 & 25.72 & -59.89& 90.43\\
$5\%$ & 20 & 28.33 &  35.24 &  35.21 & 35.34 &23.91 & 99.25\\
$6\%$ & 16.67& 40.33 &  46.70  &  46.69 & 46.74& 64.48 & 111.1\\
$7\%$ & 14.29 & 53.31 & 59.32 &  59.29 & 59.25&92.80&125.1 \\
$8\%$ & 12.50& 66.99 & 72.73  & 72.68 & 72.59 & 116.3&140.2\\
$9\%$ & 11.11& 81.23 & 86.76 & 86.75 & 86.58 & 137.5&155.9\\
$10\%$ & 10.00& 95.81 & 101.2 & 101.1  & 101.2 &157.2&172.0\\
$15\%$ & 6.67& 171.9 & 176.7 & 176.6 &176.9 &249.5 &256.1 \\
\bottomrule
\end{tabular*}
}} \end{center}\caption{\footnotesize\bf Fair fee $\alpha$ in bp ($1\mathrm{bp}=0.01\%$) of GMWDB with DB0, DB1, and DB2 death
benefits for the \emph{static case} with a quarterly withdrawal frequency
($N_w=4$) as a function of annual contractual rate $g$. ``no death" corresponds to GMWB (i.e. GMWDB with death probability set to zero).  Other parameters are $r=5\%$ and
$\sigma =20\%$.} \label{tab_feeSD1}
\end{table}

\subsubsection{Dynamic GMWDB}
 Table
 \ref{tab_feeDD1} shows results for dynamic GMWDB with DB0, DB1 and DB2 death
 benefits and dynamic GMWB calculated  using GHQC method. Here we also present
 FD method results for DB0 case. The maximum difference of fees between GHQC and FD methods is
0.4 basis point occurring at the shortest maturity, which has the
maximum magnitude in the fee. In relative terms this is less than
$0.2\%$. For other cases the difference between methods is similar, i.e. very small and we show results calculated using GHQC method only.

\begin{figure}[!h]
\begin{center}
\includegraphics[scale=0.8]{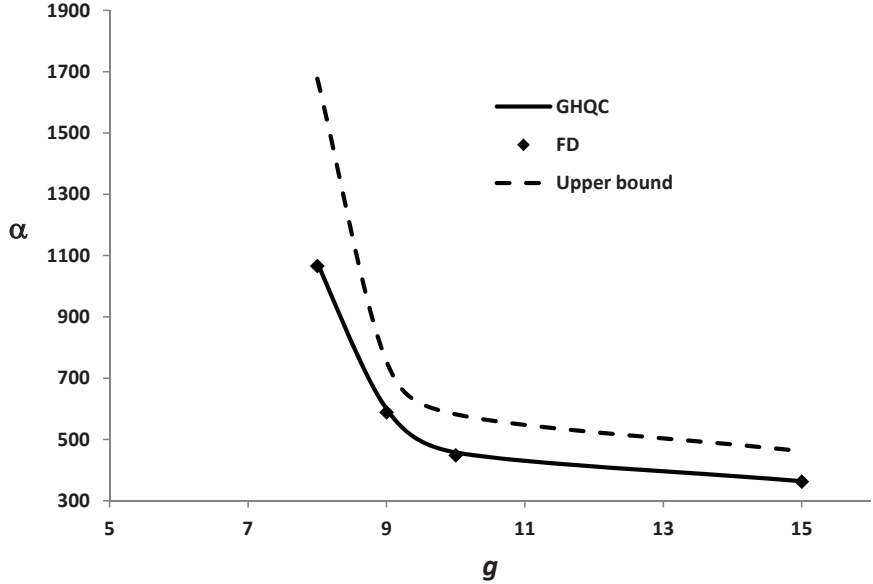}
\vspace{0cm} \caption{\footnotesize\bf Fair fee $\alpha$ in bp ($1\mathrm{bp}=0.01\%$) of GMWDB with DB2 death
benefit for the dynamic case with a quarterly withdrawal frequency
($N_w=4$). The other parameters are $r=5\%$,
$\sigma =20\%$, $\beta=10\%$. The  ``Upper bound" is the fair fee
associated with the annuity price upper estimator
calculated using equation (\ref{GMWDB_upper_eq}) based on optimal strategy if death time is known.
}\label{fig_feeDD2}
\end{center}
\end{figure}

Comparing with results for GMWB, adding
 DB0 death benefit to a dynamic GMWB only requires a maximum of 10
extra basis points in the fair fee, similar to the static case.
As easily seen from the table, adding DB1 or DB2 death benefit (returning at
least the initial premium) changes the situation dramatically. At
$g=15\%$, the shortest maturity, the fair fee of GMWDB with DB1 or
DB2  death benefit more than doubled from the value of
GMWB. Moreover, the fair fee increases rapidly as the maturity
increases, while in all the other cases so far shown in this paper
the fair fee is a decreasing function of maturity. This increase in
the fair fee is so rapid that  no solution for the fair fee exists
for $g \leq 7\%$ (i.e. for $T=1/g\geq 14.29$) for either DB1 or DB2 -- meaning that even charging
 fee at $100\%$ is still not enough to cover the risks. The last
column in the table is the upper bound
 of the fair fee corresponding to the upper estimator $Q_0^{(\mathrm{u})}\left(W(0),A(0),I_0=1\right)$ calculated using
equation (\ref{GMWDB_upper_eq}), i.e. estimator calculating GMWDB for given death time (conditional on knowing death time) and then averaging over possible death times with corresponding death probabilities. Figure \ref{fig_feeDD2} plots the
fair fee of GMWDB with DB2 as a function of the contractual withdrawal
rate $g$, showing the rapid increase of fair fee as the contractual rate
decreases or the maturity increases. Between DB1 and DB2, the
difference in the fair fee is very small, unlike the static
case where the difference in the fair fee between DB1 and DB2 is
significant. Also shown in Figure \ref{fig_feeDD2} (dashed line) is
the the upper bound of the fair fee from Table
\ref{tab_feeDD1}. The upper bound of the fee is
much higher than the fee corresponding to the
optimal withdraw strategy based on information up to the withdrawal date,
manifesting the value of `knowing the future'.

\begin{table}[!h]
\begin{center}
{\footnotesize{\begin{tabular*}{1.0\textwidth}{cccccccc} \toprule
\textbf{contractual rate} & \bf{maturity} &  \bf{GHQC} & \bf{GHQC} &   \bf{FD} & \bf{GHQC} & \bf{GHQC} & \bf{GHQC}\\
{$g$} & {$T=1/g$} & {no death} & {DB0}  & {DB0} & {DB1} & {DB2} & {DB2$^\ast$}\\
 \midrule
 $4\%$ & 25.00 &  56.09 & 66.43 & 66.51 & N/A& N/A& N/A\\
$5\%$ & 20.00 & 70.07&  77.93 &  77.95 & N/A& N/A& N/A\\
$6\%$ & 16.67 & 83.74&  90.32  &  90.29 & N/A& N/A& N/A\\
$7\%$ & 14.29 & 97.11&  102.9 &  102.8 & N/A& N/A& N/A\\
$8\%$ & 12.50 & 110.3 & 115.6  & 115.4 & 1072 & 1076   & 1677 \\
$9\%$ & 11.11 & 123.2& 128.1& 127.9&  597.6 & 599.6 & 754.6\\
$10\%$ & 10.00 & 136.0& 140.6 & 140.4&  455.9 & 457.7 & 582.2 \\
$15\%$ & 6.67 &  199.0 & 203.0 & 202.6&  362.1 & 363.8  & 461.9\\
\bottomrule
\end{tabular*}
}} \end{center} \caption{{\bf\footnotesize Fair fee $\alpha$ in bp ($1\mathrm{bp}=0.01\%$)  of GMWDB with DB0, DB1 and DB2 death
benefits for the \emph{dynamic case} with a quarterly withdrawal frequency
($N_w=4$).  The other parameters are $r=5\%$,
$\sigma =20\%$ and $\beta=10\%$. ``no death" corresponds to GMWB (GMWDB with death probability set to zero). The last column, {DB2$^\ast$}, results are fair fees corresponding to the annuity upper estimator based on the optimal strategy in the case of perfect death time forecast given by equation
(\ref{GMWDB_upper_eq}). }} \label{tab_feeDD1}
\end{table}

\begin{figure}[!h]
\begin{center}
\includegraphics[scale=0.8]{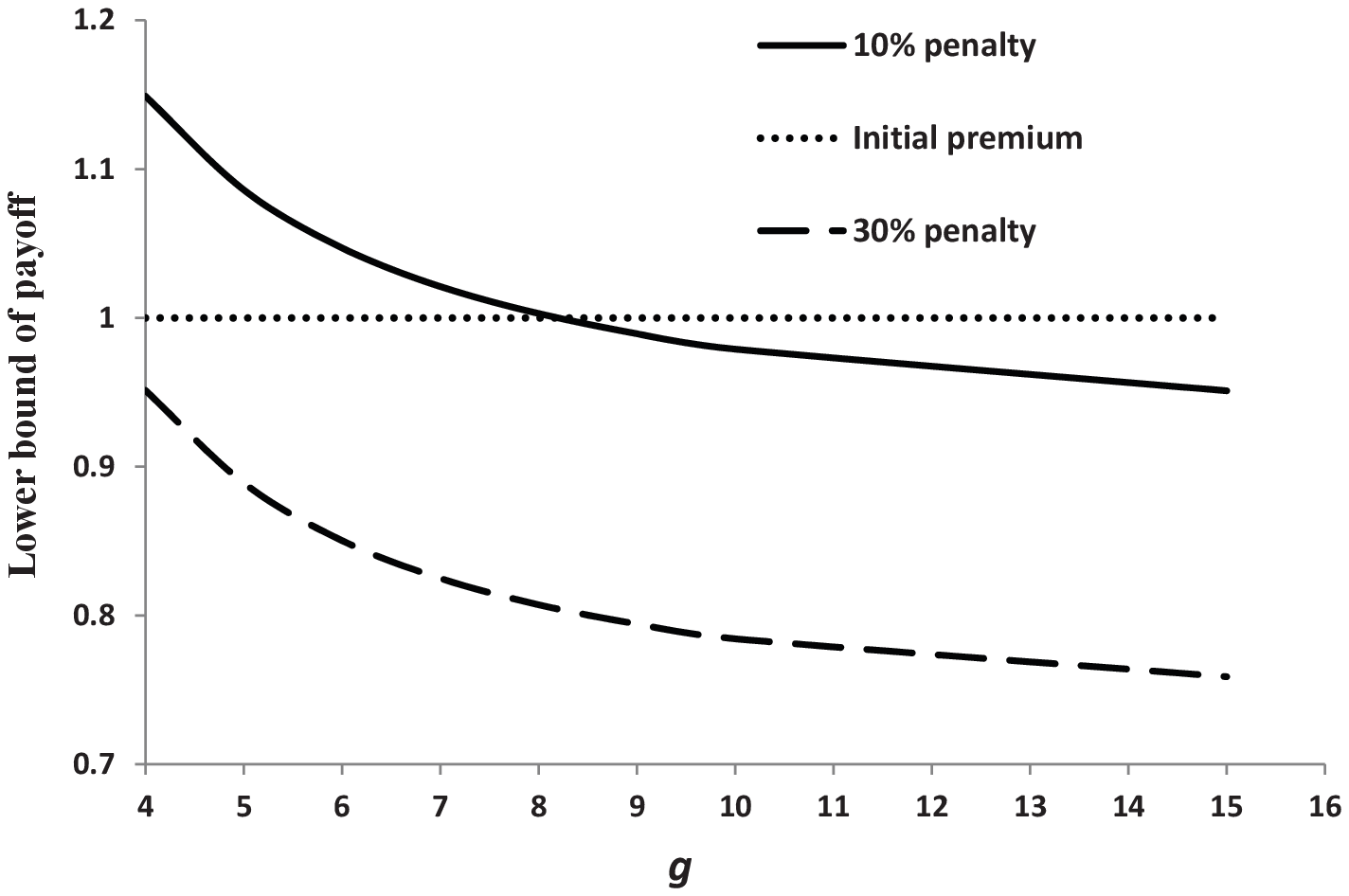}
\vspace{0cm} \caption{\footnotesize\bf Lower bounds for the payoff $P_L$ for the
simple strategy for the dynamic case with a quarterly withdrawal
frequency ($N_w=4$) and death benefit DB1 or DB2, at $\beta=10\%$
and $\beta=30\%$. The other parameters are
$r=5\%$, $\sigma =20\%$. }\label{fig_PL}
\end{center}
\end{figure}

Now let us look at the reason for the  non-existence of solution at
longer maturity
 by considering the following  simple pre-defined strategy (sufficiently good strategy but not necessarily
 optimal thus corresponding to the lower price estimator). Assume the policyholder of a dynamic GMWDB with  DB2
  withdraws all guarantee amount $A(0)$ at the first withdrawal date
 (if surviving the first time period) and waiting for the possible collection of the death benefit. Then wealth account $W(t_n)$ evolves according to (\ref{wealth_process_eq}) with $\gamma_1=A(0)$, $\gamma_n=1, n=2,\ldots,N$.
The expected present value of payoff (expectation with respect to death time $\tau$ and wealth process $W({t_n})$) received by the policyholder using the
above strategy is


$$
\begin{array}{lll}
  \mathrm{E}_{t_0}^{\bm{X}}[H_0(\bm{X},\bm{\gamma})]& = & \mathrm{E}_{t_0}^{\bm{W}}\left[B_{0,N}\max(W(T^-),0)\Pr[\tau>T]\right. \nonumber\\
  &&\left.+(1-p_1) C_1(A(0)) B_{0,1} + \sum_{i=1}^N \; p_n\max(W(0), W(t_n^-) )B_{0,n}\right] \\
 & \geq  & (1-p_1)C_1(A(0)) B_{0,1}+W(0)\sum_{i=1}^N \; p_i B_{0,i}\equiv Q_0^{(\mathrm{L})},
\end{array}
$$

\noindent where $p_n=\Pr[t_{n-1}<\tau<t_{n}]$ is the probability of death occurring during
the $n$-th payment period conditional on the policyholder is alive at
the beginning of the contract, $C_1(A(0))=G_1+(W(0)-G_1)(1-\beta)$ is the full
withdraw amount minus penalty if the policyholder survives the first
payment period to execute the strategy, and $Q_0^{(\mathrm{L})}$ denotes a lower
bound of the expected payoff for this simple strategy.
The above estimate of the payoff for the strategy
is independent of the fee charged by the insurance company - because
the fee only affects the account value, not the guarantee amount,
nor the minimum death benefit. Using the same strategy, the above
lower bound also applies to DB1. Figure \ref{fig_PL} shows the lower
bound $Q_0^{(\mathrm{L})}$ as a function of
the contractual rate $g$. Clearly, at $g \leq 7$ and $\beta=10\%$
the strategy always yields a cashflow greater than the initial
premium,
 irrespective of the
fee charged, thus explaining why no solution of fair fee exists for
$g \leq 7$. Qualitatively, GMWDB with the minimum death benefit of
returning the premium allows a rational policyholder of a long
maturity contract to get almost all the initial premium back while
 having a high probability of collecting the death benefit as well.


One remedy for the above problem of non-existence of solution is
simply increasing the penalty rate to discourage excessive
withdrawals above the contractual rate. Also shown in Figure
\ref{fig_PL} is another payoff curve in the case of penalty rate increased to
$\beta=30\%$ -- in this instance the estimated payoff is well below
the initial premium at all withdrawal rates $g\geq 4\%$ and solution
for the fair fee exists.

Also, as already mentioned in previous sections, real product design may impose expiry date for the death benefit guarantee  (e.g. at the age 70 or 75) that will reduce the cost of the death benefit guarantee itself and will help to avoid the case of non-existence of solution.

\subsubsection{Dynamic GMWDB with fixed installment fees for death benefit}
The above problem of non-existence of fair fee solution for dynamic GMWDB in some cases, can also be dealt with by a more reasonable fee structure. So far the extra cost of adding DB1 or DB2  death benefit
 is absorbed in the continuous fee $\alpha$ which is linked to the
 wealth account value $W(t)$, not to the guaranteed premium in the death
 benefit. As the fund value is reduced by each withdrawal, the charge
 base is also reduced. A better fee structure is to keep the
 continuous fee the same as the dynamic GMWB but charge the extra
 fee due to adding DB1 or DB2 death benefit upfront or in fixed
 instalments. In practice the later is usually preferred.

\begin{figure}[!h]
\begin{center}
\includegraphics[scale=0.8]{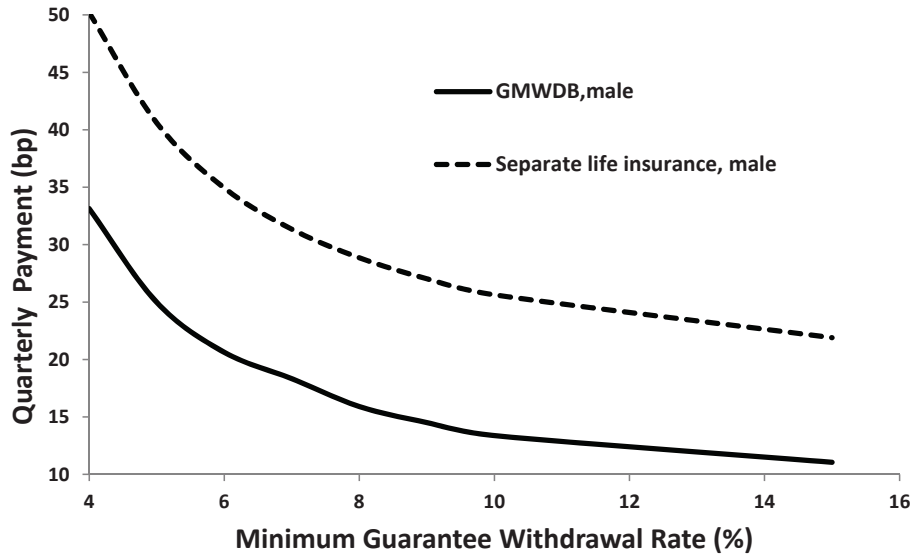}
\vspace{0cm} \caption{\footnotesize\bf Fair instalment fee $\lambda$ in bp ($1\mathrm{bp}=0.01\%$) for GMWDB with
DB2 death benefit and $\widetilde{\lambda}$ for life insurance, the inputs are the same as for results in Table \ref{tab_lambda}.
}\label{fig_lambda}
\end{center}
\end{figure}

Denote the fair fee of a dynamic GMWB (i.e. GMWDB with death probabilities set to zero) as $\alpha^\ast$ (e.g. column 3 in Table \ref{tab_feeDD1}), and the fair price of a GMWDB (with either DB1 or
 DB2)
  under the same fee
 as $Q{(\alpha^\ast)}\equiv Q_0(A(0),W(0),I_0=1)$, then obviously
 $Q{(\alpha^\ast)}>W(0)$ and the fair upfront fee is  the difference $\triangle=Q{(\alpha^\ast)}-W(0)$.
 To work out the constant instalment amount $\lambda$ , the probability of
 death has to be considered. Here we assume each instalment is paid
 at the beginning of each withdrawal period and no further payment is
 required upon death. Conditional on death occurring during the
 period $(t_{n-1}, t_n],\; 1\leq n \leq N$, i.e. $t_{n-1} < \tau \leq
 t_n$, the total sum of instalments received by the issuer of GMWDB is given by
$\Lambda_n = \lambda\sum_{j=1}^n B_{0,j-1}$, and the expected
total sum of instalments is then
$$\overline{\Lambda}=p_S \Lambda_N+\sum_{n=1}^N \Lambda_n p_n,$$
where $p_S=\Pr[\tau>T]$ is the probability of surviving beyond maturity $T$, and
$p_n=\Pr[t_{n-1}<\tau\le t_n]$ is the probability of death occurring during the
 period $(t_{n-1}, t_n]$. Equating $\overline{\Lambda}$ with the
 upfront fee $\triangle=Q{(\alpha^\ast)}-W(0)$, we obtain the fair
 instalment fee $\lambda$
 \begin{equation}\label{eq_lambda}
  \lambda =\frac{ Q{(\alpha^\ast)}-W(0) }{p_S \rho_N + \sum_{n=1}^N \rho_n p_n}, \;\; \rho_n=\sum_{j=1}^n B_{0,j-1}.
\end{equation}
Using (\ref{eq_lambda}) the calculation of instalment $\lambda$ for
GMWDB  given the fair fee for GMWB is straightforward. Table
\ref{tab_lambda} shows the values of $\lambda$ in terms of basis
points of the initial premium for the same cases as in Table
\ref{tab_feeDD1}. It is interesting and useful to compare the extra
instalment $\lambda$ due to adding DB1 or DB2 death benefit to GMWB
against the instalment $\widetilde{\lambda}$ of a separate life
insurance with the same term, instalment frequency  and the insured
amount equal to the initial premium of GMWDB, $W(0)$. In the case of
such a life insurance, the instalment $\widetilde{\lambda}$ is given
by
\begin{equation}\label{eq_lambdat}
  \widetilde{\lambda} =\frac{ W(0)\sum_{n=1}^N p_n B_{0,n}}{p_S \rho_N + \sum_{n=1}^N \rho_n p_n},
\end{equation}
where $ W(0)\sum_{n=1}^N p_n B_{0,n}$ is the expected payoff of the
life insurance. Table \ref{tab_lambda} also shows the values of
$\widetilde{\lambda}$ in comparison with the values of $\lambda$.
Clearly the instalment fee of buying a separate life insurance is
higher than the extra instalment fee by adding DB1 (life insurance)
or DB2 (GMDB payoff) on top of GMWB,
 making the combined contract GMWDB an attractive product. The life insurance fee is about $51\%$ higher than
 the extra fee of GMWDB with DB2 at the longest maturity corresponding to $g=4\%$, and this difference increases
 to $98\%$ at the shortest maturity corresponding to $g=15\%$.
 What is
 more, upon death the GMWDB contract with DB2 returns the maximum between
 account value $W(t_d)$ (where $t_d$ is first time slice after the death) and the initial premium $W(0)$, while the life insurance
 only returns the insured sum $W(0)$. The
 extra instalment required by  DB1 is only very slight smaller than that of DB2.
Figure \ref{fig_lambda} shows the curve of
 $\lambda$ of DB2 in comparison with $\widetilde{\lambda}$ for the male population. The
 curve for DB1 (not shown) is hardly distinguishable from those of DB2 in Figure
 \ref{fig_lambda}.

 Instead of buying a separate
 life insurance, another alternative is to buy a GMDB contract in
 addition to the GMWB contract, in this case it is very expensive
 because GMDB requires an additional  premium of $W(0)$, while the life
 insurance does not, nor does the combined GMWDB with either DB1 or DB2 (only a single premium is required).
  So overall it is better and cheaper to buy a single combined
 GMWDB contract with either DB1 or DB2 death benefit.

Table \ref{tab_female} is the female counterpart of Table
\ref{tab_lambda}. As expected, both  the extra instalment fee of
 the GMWDB on top of the GMWB and the separate life insurance for the female population is
much cheaper than for the male population. Depending the maturity,
there is also some significant saving on
 fees of about $40\%$ to $50\%$ for the female population, if the GMWDB is
purchased instead of buying a separate life insurance on top of
GMWB, although the saving is not as great as for the male
population.

\begin{table}[!h]
\begin{center}
{\footnotesize{\begin{tabular*}{0.75\textwidth}{ccccc} \toprule
contractual rate, g & maturity $T=1/g$ &   $\lambda$ (DB1)  &  $\lambda$ (DB2)  &  $\widetilde{\lambda}$   \\
 \midrule
 $4\%$ & 25.00 &32.05 &33.13 &  50.20   \\
$5\%$ & 20.00 & 24.39 &24.97 &  40.61 \\
$6\%$ & 16.67 & 20.23 &20.63  &  34.94 \\
$7\%$ & 14.29 & 18.01 &18.33 &  31.35 \\
$8\%$ & 12.50 & 15.59 &15.91 & 28.86  \\
$9\%$ & 11.11 & 14.18 &14.51& 27.03  \\
$10\%$ & 10.00 &13.03 &13.37 & 25.63 \\
$15\%$ & 6.67 & 10.70 &11.05 &  21.90 \\
\bottomrule
\end{tabular*}
}} \end{center}\caption{\footnotesize\bf Fair instalment fee $\lambda$ in bp ($1\mathrm{bp}=0.01\%$) of GMWDB with DB1 or DB2 death benefit for the \emph{dynamic case} with a quarterly
withdrawal frequency ($N_w=4$), male population. This instalment is
on top of the continuous fair fee of a standard GMWB contract. The
other parameters are $r=5\%$, $\sigma =20\%$,
$\beta=10\%$. $\widetilde{\lambda}$ is the instalment fee for a
separate life insurance of the same term with the insured sum equal
to $W(0)$, the initial premium of GMWDB.} \label{tab_lambda}
\end{table}

\begin{table}[!h]
\begin{center}
{\footnotesize{\begin{tabular*}{0.7\textwidth}{ccccc} \toprule
contractual rate, g & maturity $T=1/g$ &  $\lambda$ (DB1)  &  $\lambda$ (DB2)  & $\widetilde{\lambda}$   \\
 \midrule
 $4\%$ & 25.00 &18.90 &20.14 &  32.55   \\
$5\%$ & 20.00 & 13.34 &14.14 &  24.85 \\
$6\%$ & 16.67 & 10.84 &11.46  & 20.96 \\
$7\%$ & 14.29 & 9.67 &10.23 &  18.62 \\
$8\%$ & 12.50 & 8.23 &8.82 & 17.01 \\
$9\%$ & 11.11 & 7.49 &8.06& 15.82 \\
$10\%$ & 10.00 & 6.89 &7.43 & 14.97 \\
$15\%$ & 6.67 & 6.02 &6.35 &  12.86 \\
\bottomrule
\end{tabular*}
}} \end{center}\caption{\footnotesize\bf Fair instalment fee $\lambda$ in bp ($1\mathrm{bp}=0.01\%$) of GMWDB with
DB1 or DB2 death benefit for the dynamic case with a quarterly
withdrawal frequency ($N_w=4$), female population. This instalment
is on top of the continuous fair fee of a standard GMWB contract.
The other parameters are the same as in Table
\ref{tab_lambda}. $\widetilde{\lambda}$ is the instalment fee for a
separate life insurance of the same term with the insured sum equal
to $W(0)$, the initial premium of GMWDB.} \label{tab_female}
\end{table}

\break

\section{Conclusion}\label{conclusion_sec}
 In this paper we have presented a numerical valuation of a variable
annuity with GMWDB (i.e. with combined GMWB and GMDB features) under both passive (static) and optimal (dynamic) policyholder
behaviors. Essentially this contract simultaneously deals financial
risk,  mortality risk and human behavior in
terms of decision under uncertainty. In the case of optimal policyholder behaviour, we formulate and solve the valuation as a stochastic control problem for controlled Markov process, i.e. policyholder performs optimal withdrawals based on account information available at withdrawal date.
It is important to note that pricing dynamic GMWDB contract for given death time (i.e. conditional on knowing death time) and then averaging over different death times according to death probabilities will lead to higher price than dynamic GMWDB where decisions are based on information available at withdrawal date.

We presented a series of numerical results  for  GMWDB with different death benefit types and showed that, in the case of optimal policyholder behavior,  adding the premium for the minimum death benefit on top of the
  GMWB can be problematic for contracts with long maturities if the
   continuous fee structure is kept, which is ordinarily assumed for a GMWB
   contract.
  In fact for some long maturities we observed
 that  the fee cannot be charged as any proportion of the wealth account value --  there is no solution to match the initial premium with the fair annuity price. To avoid this problem, the product design may impose expiry for the death benefit guarantee, e.g. at the age of 70 years.
    On the other  hand, the extra fee due to adding the death benefit can be
  charged upfront or in periodic instalment of fixed amount and it is cheaper than buying a separate life insurance.

\section{Acknowledgement}
In conclusion, we are grateful to Juri Hinz for useful discussions of the problem. This research was supported by the CSIRO-Monash
Superannuation Research Cluster, a collaboration among CSIRO, Monash University, Griffith University, the University of Western Australia, the University of Warwick, and stakeholders of the retirement system in the interest of better outcomes for all.


\bibliographystyle{authordate1} 
{\footnotesize
\bibliography{bibliography}
}
\newpage

\appendix
\section{Australian Life Table}
\begin{table}[!h]
\begin{center}
{\footnotesize{
\begin{tabular*}{0.6\textwidth}{ccc} \toprule
Age & Number of males  &  Number of females   \\
& surviving to exact age & surviving to exact age \\
 \midrule
 0 & 100,000 & 100,000 \\
 ... & ... & ...\\
 60  &91,305 & 94,817 \\
61   &90,684 & 94,434 \\
62 & 90,010 & 94,019\\
63 & 89,276&  93,566\\
64 &88,475& 93,073\\
65& 87,601& 92,535\\
66&  86,646 &91,948\\
67&  85,603& 91,306\\
68& 84,463& 90,604\\
69& 83,219& 89,838\\
70& 81,863& 88,997\\
71& 80,390& 88,053\\
72& 78,791& 86,997\\
73&  77,061& 85,836\\
74& 75,191& 84,571\\
75& 73,169& 83,192\\
76& 70,982& 81,683\\
77& 68,614& 80,026\\
78& 66,053& 78,196\\
79& 63,287& 76,168\\
80& 60,308& 73,916\\
81& 57,115& 71,412\\
82& 53,710& 68,636\\
83& 50,109& 65,568\\
84& 46,333& 62,200\\
85& 42,415& 58,532\\
 ... & ... & ...\\
\bottomrule
\end{tabular*}
}} \end{center}\caption{\footnotesize\bf Australian Life Table. Source: Australian
Bureau of Statistics (3302055001DO001-20092011
Life Tables, States, Territories and Australia 2009-2011, released at
11:30 am Canberra time 8 November 2012, available on http://www.abs.gov.au/).
}\label{tab_life}
\end{table}

\end{document}